\documentclass[traditabstract,preprint]{aa}
\usepackage{epsfig,graphicx,natbib}
\usepackage{longtable}
\usepackage{amssymb}

\def\PKS1830{\hbox{PKS\,1830$-$211}}

\begin{document}

\title{Calibration of mixed-polarization interferometric observations}

\subtitle{Tools for the reduction of interferometric data from elements with linear and circular polarization receivers}

\author{I. Mart\'i-Vidal\inst{1} \and 
        A. Roy\inst{2} \and
        J. Conway\inst{1} \and
        A. J. Zensus\inst{2} 
}

\offprints{I. Mart\'i-Vidal \\ \email{mivan@chalmers.se}}
\institute{Department of Earth and Space Sciences, Chalmers University of Technology, 
           Onsala Space Observatory, SE-43992 Onsala, Sweden  \and
           Max-Planck-Institut f\"ur Radioastronomie, Auf dem H\"ugel 69, 
           D-53121 Bonn, Germany
}

\date{Accepted for publication in A\&A}
\titlerunning{Mixed-polarization interferometric observations}
\authorrunning{Mart\'i-Vidal et al. (2015)}

\abstract{
Heterodyne receivers register the sky signal on either a circular polarization basis (where it is split into left-hand and right-hand circular polarization) or  a linear polarization basis (where it is split into horizontal and vertical linear polarization). We study the problem of interferometric observations performed with telescopes that observe on different polarization bases, hence producing visibilities that we call  ``mixed basis'' (i.e., linear in one telescope and circular in the other). We present novel algorithms for the proper calibration and treatment of such interferometric observations and test our algorithms with both simulations and real data.
The use of our algorithms will be important for the optimum calibration of forthcoming observations with the Atacama Large mm/submm Array (ALMA) in very-long-baseline interferometry (VLBI) mode. Our algorithms will also allow us to optimally calibrate future VLBI observations at very high data rates (i.e., wide bandwidths), where linear-polarization feeds will be preferable at some stations, to overcome the polarimetric limitations due to the use of quarter-wave plates.

}


\maketitle

\section{Introduction}

Most heterodyne receivers record  signals on a genuine linear basis. The receiver's front end can be understood as two orthogonal dipoles, one horizontal (the X-axis) and one vertical (the Y-axis), in the frame of the antenna mount (although there could also be a generic rotation of the XY plane on a third antenna axis, parallel to the source direction). The Atacama large mm/submm array (ALMA) is a good example of an interferometer that uses linear-feed polarization receivers \citep[e.g., ][]{lamb}. 

However, VLBI observations involve baselines that are so long that the parallactic angle $\psi$ of two antennas pointing to the same source can be quite different (assuming that the mounts of the antennas are alt-azimuthal). This orientation issue in VLBI can cause many problems if the signals are recorded on a linear polarization basis since the gain and polarization solutions are coupled in this case \citep{Sault}. Hence, the amplitude calibration must be known before any $\psi$ correction can be applied. Given that it is desirable to apply the $\psi$ correction as early as possible to prevent its aggregation into more critical phase-calibration steps, such as global fringe-fitting \citep{schwab} and phase referencing \citep{beasley}, the use of linear-polarization feeds in VLBI does not seem to be a practical choice. 

Nevertheless, if the signals are recorded on a circular polarization basis (i.e., right-hand, RCP, and left-hand, LCP, circular polarizations), the gain and polarization solutions are decoupled through the variation of $\psi$ \citep{Sault}.
In this R/L basis, the $\psi$ correction becomes (to a first-order approximation) a mere phase addition to RCP and LCP so there is no need to have either similar $T_{sys}$ or known amplitude correction factors in the initial phase calibration. The $\psi$ correction on this basis commutes with the gain calibration (both in amplitude and phase). In addition, if either of the channels (RCP or LCP) fails to be recorded (or  is of bad quality), it is still possible to use the other channel for the science analysis (as long as the observed sources do not have circular polarization). This is not possible if the data are being observed on the X/Y basis  since, in this case, both channels will always be needed for the $\psi$ correction.

Circular polarization receivers can be built by just adding a quarter-wave plate (or an equivalent device) at the  front end of the receiver so that the incoming circular polarization is converted into linear and then detected by the dipoles \citep[e.g., ][]{marrone}. There are, however, some disadvantages to the use of these polarizers. The purity of the polarization is usually lower (there is more polarization leakage between the channels) and the effective bandwidth is also narrower. Quarter-wave plates are designed to be optimum at a given frequency, but their performance degrades as the signal frequency departs from that used in the quarter-wave plate design. These disadvantages make linear polarizers the optimum choice for interferometers with wide-band receivers and high polarization purity, like ALMA and the SKA. For instance, most of the stations of the global mm-VLBI array (GMVA), which use circular polarizers at 86\,GHz, suffer from polarization leakages as high as 2$-$6\% \citep{GMVA}; the reported leakage of the circular polarizers at the submillimeter array (SMA) can also reach values up to 4\% \citep{marrone}.  In contrast, the leakage from the linear polarizers at ALMA reach amplitudes of $\sim$1\% \citep[e.g., ][]{lamb}.

It is possible to find the situation where two stations with different types of receivers (linear and circular) have to become part of the same interferometer. A good example  is the phased ALMA used in VLBI mode. As we have already noted, ALMA uses receivers that record the signal on a linear (X/Y) basis, whereas VLBI stations mostly record the signals on a circular (R/L) basis. As a consequence, when the phased ALMA is used as a VLBI station, the visibilities between ALMA and the other VLBI antennas will be correlated on a mixed polarization basis (i.e., linear for ALMA, circular for the other stations). These visibilities have to be properly calibrated and converted to a pure circular basis before any further analysis can be performed. 

An alternative approach to the mixed-polarization correlation would be to convert the ALMA data streams to an R/L basis before the correlation. However, neither the phase (and delay) offset(s) between X and Y nor the difference in the X/Y amplitude gains (and their time evolution) are known at the correlation stage. All these quantities are needed for an optimal polarization conversion. Basically, a gain offset between X and Y translates into a combination of gain-like and leakage-like effects in R/L, whose calibration is more complicated than that of a simple phase/amplitude gain \citep[the leakage and phase terms interact in the fit in a non-trivial way, dependent on source structure and parallactic angle coverage, e.g., ][]{Cotton93,Sault}. Hence, a blind X/Y to R/L conversion at the correlation stage, with no preliminary phase and amplitude corrections, will be far from optimal.  

In this paper, we study the problem of how to calibrate and deal with visibilities using mixed polarization. In the next section, we describe the algorithms for the calibration of mixed-polarization visibilities related to a phased array with linear-feed receivers (e.g., ALMA) and related to a single station with linear-feed receivers (e.g., a VLBI station with no quarter-wave plate). In Section \ref{PolConvSec}, we present our software, where these algorithms are implemented. In Section \ref{Simuls}, we test the performance of our algorithms with realistic simulations. In Sect. \ref{RealData}, we present the results of the calibration of real VLBI observations on mixed-polarization basis. In Section \ref{Summary}, we summarize our conclusions.

\section{RIME formalism for polarimetry}
\label{TheorySec0}

We will make use of the formalism of the Radio Interferometer Measurement Equation, RIME \citep[e.g.,][]{Hamaker}. The RIME provides an abstraction of polarization basis (e.g., circular or linear) in a convenient mathematical framework in which the basis choice does not propagate explicitly in the full algebra for the interferometer response \citep{Hamaker,Sault}. \cite{smirnov2011a} makes use of this abstraction for a convenient formulation of the calibration problem on different polarization bases, and the problem of calibration/conversion from one basis to the other in a direct and elegant way. 

In the following lines, we briefly summarize some of the results by \cite{smirnov2011a,smirnov2011b} that will be used in this paper. We refer the reader to the original publications by Smirnov, for a more detailed discussion. 
We will use the the $\odot$ symbol to refer to the circular polarization basis and the $+$ symbol to refer to the linear polarization.

The parameters $x$ and $y$ are two orthonormal vectors on the plane perpendicular to the wave propagation;   $e_x$ and $e_y$ are the electric-field components of the source signal in the $x$ and $y$ directions, respectively.  
From $e_x$ and $e_y$ we can compute the electric field on a circular polarization basis. The conversion is straightforward,

\begin{equation}
e_r = \frac{1}{\sqrt{2}}(e_x - j\,e_y)~~~;~~~e_l = \frac{1}{\sqrt{2}}(e_x + j\,e_y),
\label{Lin2Circ}
\end{equation}

\noindent where $r$ and $l$ represent the right-hand and left-hand circular polarizations, respectively, and $j$ is the imaginary unit.
For a given brightness distribution on the sky, we can define the cross-correlation matrix in circular polarization basis, $E_{\odot \odot}$, for a given baseline and time, in the form

\begin{equation}
E_{\odot \odot} = \left(
\begin{array}{cc}
< e_r e'^*_r > & < e_r e'^*_l > \\ 
< e_l e'^*_r > & < e_l e'^*_l >
\end{array} \right),
\end{equation}

\noindent where $e$ is the electric field at the first antenna and $e'$ is the electric field at the second antenna. This is the equivalent to the left-hand side of Eq. 7 in \cite{smirnov2011a}, but for circular-feed receivers. For small fields of view, and if direction-dependent effects in the antenna gains are neglected, this matrix is related to the Fourier transform of the brightness matrix defined in \cite{smirnov2011a}. If the first antenna of the baseline is sensitive to linear polarization, the cross-correlation matrix (i.e., $E_{+ \odot}$) will  instead be

\begin{equation}
E_{+ \odot} = \left(
\begin{array}{cc}
< e_x e'^*_r > & < e_x e'^*_l > \\ 
< e_y e'^*_r > & < e_y e'^*_l >
\end{array} \right).
\end{equation}

We can define the matrix $E_{\odot +}$ (i.e., linear polarization for the second antenna of the baseline and circular polarization for the first one) in a similar way. As  is shown in \cite{smirnov2011a} (see his Sect. 6.3), from Eq. \ref{Lin2Circ} it follows that

\begin{equation}
\left(
\begin{array}{c} 
e_r \\ e_l \end{array} 
\right) = C_{\odot +} 
\left( \begin{array}{c} e_x \\ e_y \end{array} 
\right),
\label{L2CEq}
\end{equation}

where 

\begin{equation}
C_{\odot +} = \frac{1}{\sqrt{2}}\left(
\begin{array}{cc}
 1 & -i \\ 
 1 & +i
\end{array} \right) ~~~ \mathrm{and} ~~~ C_{+ \odot} = C^H_{\odot +}.
\end{equation}

The matrices $C_{\odot +}$ and $C_{+ \odot}$ respectively convert polarizations from linear to circular and from circular to linear. In \cite{smirnov2011a} these matrices are called $H$ (for {\em hybrid}); here we  call them $C$ (for {\em conversion}).

\section{Calibration of mixed-polarization visibilities}
\label{TheorySec}

The matrices $C_{\odot +}$ and $C_{+ \odot}$ can convert the cross-correlation matrix given on the basis of mixed polarizations (i.e., circular-linear or linear-circular) into a matrix given on the basis of pure circular polarization  by applying it to just one side of the RIME. Hence, we can recover the cross-correlation matrix in circular-circular polarization directly from the matrix given in mixed polarizations by applying a simple matrix product.

If we cross-correlate the voltages registered at the receivers of two antennas sensitive to circular polarization, the resulting visibility matrix is

\begin{equation}
V_{\odot \odot} = \left(
\begin{array}{cc}
< v_r v'^*_r > & < v_r v'^*_l > \\ 
< v_l v'^*_r > & < v_l v'^*_l >
\end{array} \right),
\end{equation}

\noindent where $V_{\odot \odot}$ is the visibility matrix with circular polarization in both antennas, $v$ is the voltage in the first antenna, and $v'$ is the voltage of the second antenna. If the first antenna of the baseline is sensitive to linear polarization, the visibility matrix (i.e., $V_{+ \odot}$) will instead be

\begin{equation}
V_{+ \odot} = \left(
\begin{array}{cc}
< v_x v'^*_r > & < v_x v'^*_l > \\ 
< v_y v'^*_r > & < v_y v'^*_l >
\end{array} \right).
\end{equation}

We can define  the matrix $V_{\odot +}$ in a similar way.  
If $J$ and $J'$ are the Jones matrices that fully calibrate the first and second antennas, respectively, of the baseline, then, using Eq. (8) in \cite{smirnov2011a}, we can write 

\begin{equation}
V_{+ \odot} = J_{+} E_{+ \odot} J_{\odot}'^H 
\label{Kk1}
\end{equation}

and

\begin{equation}
V_{\odot +} = J_{\odot} E_{\odot +} J_{+}'^H.
\label{Kk2}
\end{equation}

\noindent where $J_{+}$ and $J'_{+}$ are the Jones matrices in the linear polarization basis and $J_{\odot}$ and $J'_{\odot}$ are the same matrices in the circular polarization basis.
Applying Eq. \ref{L2CEq}, and keeping in mind that $(AB)^{-1} = B^{-1}A^{-1}$, we have

$$ E_{\odot \odot} = C_{\odot +} J^{-1}_{+} V_{+ \odot} (J'^{-1}_{\odot})^{H} $$

and

$$ E_{\odot \odot} = J^{-1}_{\odot} V_{\odot +} (J'^{H}_{+})^{-1} C_{+ \odot}. $$

These equations allow us to relate any visibility matrix given on a mixed-polarization basis with the cross-correlation matrix given in a pure circular-circular basis.
We note that if the product $C_{\odot +} J^{-1}_{+} V_{+ \odot} (J'^{-1}_{\odot})^{H}$ is applied to each integration time of the visibilities, there is no loss of signal coherence in the conversion.

\subsection{Phased array with linear feeds}

We let $v^a_k$ be the voltage registered by an antenna $a$ at time $t_k$ in a circular polarization basis. Antenna $a$ may be, in our case, a stand-alone VLBI station with a circular-polarization feed. We let $v^i_k$ be the voltage at time $t_k$ registered by the $i$-th antenna of a phased array (which is  named $b$) in a linear polarization basis. Since the phased signal of $b$ at time $t_k$ is the addition of the voltages $v^i_k$ of all the $k$ phased antennas, the corresponding element of the visibility matrix $V(a,b)$ \citep{smirnov2011a} between the stations $a$ and $b$ is

\begin{equation}
V_{\odot +}(a,b) = \left< v^a v^b\right> = \left< v^a \sum_i{v^i}\right> = \sum_i{\left<v^a v^i \right>}. 
\label{Lab1}
\end{equation}

\noindent That is, given that the correlation operator is linear, the visibility is equal to the sum of the correlations between antenna $a$ and each individual antenna of the phased array. Taking into account all the polarization products, we can build the full visibility matrix in the same way, i.e.,

\begin{equation}
V_{\odot +}(a,b) = \sum_i{V_{\odot +}(a,i)}.
\label{Lab2}
\end{equation}

In the absence of noise, and if the signals are perfectly calibrated, all the matrices $V_{\odot +}(a,i)$ should be numerically equal (as long as the synthesized VLBI field of view is much smaller than the synthesized resolution of the phased array).

\subsection{Effect of different antenna gains in the phased elements}

The different antennas of the phased array are affected by slightly different gains, bandpass responses, and polarization leakage (D-term factors). The observed visibility matrix $V_{\odot +}^{obs}(a,b)$ is then  related to the perfectly-calibrated visibility matrix $V_{\odot +}(a,b)$ by the equation

\begin{equation}
V_{\odot +}^{obs}(a,b) = V_{\odot +}(a,b)\,J_{+}^b,
\label{calEq1}
\end{equation}

\noindent where the calibration Jones matrix $J_{+}^b$ can be a function of frequency and time, and is related to all the calibration matrices (i.e., gain, bandpass, D-terms, etc.) of all the antennas in the phased array. Thanks to the linearity of the algebra, we can write

\begin{equation}
V_{\odot +}^{obs}(a,b) = \sum_i{V_{\odot +}(a,i)\,J_{+}^i},
\label{calEq2}
\end{equation}

\noindent where antenna $i$ is corrupted by an unknown overall gain $J_{+}^i$. 
Given that the uncorrupted visibility $V_{\odot +}(a,i)$ is assumed to be noise-free, perfectly calibrated, and independent of $i$, we can put $V_{\odot +}(a,i)$ out of the sum and write (see Eq. \ref{calEq1})

\begin{equation}
J_{+}^b = \left<J_{+}^i\right>,
\label{KavgEq}
\end{equation}

\noindent where $\left<...\right>$ is the averaging operator over the antennas of the phased array. Our objective is to calculate the $J_{+}^b$ matrix of Eq. \ref{calEq1} using all the observables available. 
We let $B^i_x$ and $B^i_y$ be the bandpass gains of the $i$-th antenna in the phased array for polarization X and Y, respectively. We let $G^i_x$ and $G^i_y$ be the phase and amplitude gains for the same antennas and polarizations. We also let $\alpha_i = \exp{(j\Delta\phi_i)}$ be the relative phase-gain between X and Y for the same antenna, and $D^i_x$ and $D^i_y$ be the D-terms. 
The exact $J_{+}^b$ matrix for the full phased array becomes

\begin{equation}
J_{+}^b = \left( \begin{array}{cc}
\left<B_x\,G_x\right> & \left<D_x\,B_x\,G_x\right> \\
\left<D_y\,B_y\,G_y\,\alpha\right> & \left<B_y\,G_y\,\alpha\right> 
\end{array} \right).
\label{Lab4}
\end{equation}

The calibration matrix to be multiplied by the observed visibility matrix is  the inverse of $J_{+}^b$ (see Eqs. \ref{Kk1},  \ref{Kk2}, and \ref{calEq1}).

\subsection{Single station with linear feeds}
\label{SingleStat} 

If the station with a linear-feed receiver is a single dish, there is no standard way to compute the Jones matrices that calibrate/convert the mixed-polarization visibilities. In these cases, there are no linear-linear cross-correlations from which the gains of the linear-feed receivers can be derived as we do in Eq. \ref{KavgEq}.  
Although the conversion with the hybrid matrix $C$ is still possible, not correcting for the X and Y gains (and phase offsets) beforehand may lead to strong time-dependent leakage-like effects in the circular-circular visibilities after the polarization conversion. 

Fortunately, we can still use the information encoded in the mixed-polarization visibilities to perform a proper calibration. If $V_{rr}$ and $V_{ll}$ are the visibilities for the parallel-hand circular-circular correlations, respectively  RR and LL,  then $V_{rr}/V_{ll} = 1$ regardless of the source structure (and provided that the source is not circularly polarized). We now suppose that the first antenna of the baseline observes with a linear-feed receiver. Each polarization channel, X and Y, is affected by different gains, $B_x\times G_x$ and $B_y\times G_y$. However, {\em if the polarization leakage is negligible} the only quantity that is important for the calibration before the polarization conversion is the ratio of gains (i.e., the cross-polarization gain), $B_x\times G_x/(B_y\times G_y) = G_{x/y}$, which can  fortunately be derived easily. If we write $V_{rr}/V_{ll}$ in terms of the mixed-polarization visibilities and the gain ratios, we can define a norm $\chi^2$ whose minimum  determines the gain ratios,

\begin{equation}
\chi^2 = \sum_k{\omega_k\left[ \frac{V^k_{xr}G^{-1}_{x/y} - i\,V^k_{yr}}{V^k_{xl}G^{-1}_{x/y} + i\,V^k_{yl}}(G^*_{k,R/L})^{-1} - 1\right]^2} + \chi^2_{\odot\odot},
\label{ChiEq}
\end{equation}

\noindent where $\omega_k$ is the weight of the $k$-th visibility matrix $V^k$ and $G_{k,R/L}$ is the ratio of gains for the second antenna (with a circular-feed receiver) of the baseline of that visibility. The term $\chi^2_{\odot\odot}$ is related to the baselines that only involve antennas with circular-feed receivers,

\begin{equation}
\chi^2_{\odot\odot} = \sum_{k'}{\omega_{k'}\left[ \frac{V^{k'}_{rr}}{V^{k'}_{ll}}(G_{a_{k'},R/L})^{-1}(G^*_{b_{k'},R/L})^{-1} - 1\right]^2},
\label{ChiEqCirc}
\end{equation}

\noindent where $a_{k'}$ and $b_{k'}$ are the first and second antennas in the baseline of visibility $V^{k'}$. In this least-squares minimization, one of the antennas with circular-feed receivers has to be chosen to have a zero cross-polarization phase. This convention does not affect the calibration of the antenna with the linear-feed receiver.
 An advantage of this equation is that $G_{x/y}$ and $G_{k,R/L}$ are stable with time so we can apply long integration times to derive $\chi^2$ and use this approach even with weak sources. Once the $\chi^2$ is minimized as a function of $G_{x/y}$ and $G_{k,R/L}$, we can calibrate and convert the mixed-polarization visibilities with the equation

\begin{equation}
V^k_{\odot \odot} = C_{\odot +} \left( \begin{array}{cc} G^{-1}_{x/y} & 0 \\ 0 & 1 \end{array} \right) V^k_{+ \odot} .
\end{equation}
  
This approach is conceptually similar to self-calibration \citep{SELFCAL}, although the gain solutions are, in this case, independent of the inherent source structure if the source is not circularly polarized.
We note that this algorithm can also be useful in some cases of observations involving a phased array with linear-feed receivers. For instance, if the cross-correlations among the antennas of the phased array are not computed in full-polarization mode, or if the parallactic angle coverage of the calibrators is not large, it may not be possible to calibrate the phase offset among the X and Y signals of the reference antenna in the phased array. In such cases the data can still be calibrated, although with no D-term corrections. The approach to follow is to use Eqs. \ref{calEq1} and \ref{calEq2} (providing no phase-offset information to the Jones matrix) and to then derive the X$-$Y phase offset from the mixed-polarization visibilities using Eq. \ref{ChiEq}.

\section{Implementing the calibration/conversion of mixed-polarization visibilities}
\label{PolConvSec}

We have developed {\em PolConvert}, the calibration/conversion software that will be used in mm-VLBI observations with the phased ALMA. The program applies the calibration and conversion equations given in the previous section for a phased array with linear-feed receivers. It reads the VLBI data in standard FITS-IDI format\footnote{\texttt{http://fits.gsfc.nasa.gov/registry/fitsidi.html}}; identifies the antenna(s) with linear feeds used in the observations; and converts the visibilities to a pure circular basis, saving the results in new FITS-IDI files. In addition, the program can optionally read calibration tables from the Common Astronomy Software Applications (CASA) of the National Radio Astronomy Observatory (NRAO)\footnote{\texttt{http://casa.nrao.edu/}}, compute the Jones matrix (or matrices) for the phased array (Eq. \ref{Lab4}), and correct the VLBI visibilities (Eq. \ref{calEq1}) before the polarization conversion. Since the phased ALMA will provide the visibilities among ALMA antennas in addition to the phased signal for VLBI, the procedure to calibrate the VLBI observations will be to use the ALMA-only visibilities to derive the $J_i$ calibration matrices of all ALMA antennas (Eq. \ref{calEq2}) and to provide these matrices to {\em PolConvert} to compute the total matrix (Eq. \ref{Lab4}).   
The CASA tables currently supported in PolConvert are \citep[see][for a definition]{smirnov2011a}: G Jones (i.e., gain and X-Y phase), B Jones (i.e., bandpass), K Jones (i.e., cross delay), and D Jones (i.e., polarization leakage). As noted above, all these tables are optional for the calibration applied before the conversion. 

Several different linear-polarization VLBI stations can be corrected simultaneously, each   with its own set of CASA calibration tables. However, if the station is a single dish, no tables can be provided (unless the user knows the gain ratios between the polarization channels of the antenna receivers), so no calibration is performed before the conversion. To optimize the polarization conversion of single-dish antennas with linear receivers, we have implemented the algorithm described in Sect. \ref{SingleStat} in a script called {\em PolConvertST} to be used in the Astronomical Image Processing (AIPS) environment of NRAO \footnote{\texttt{http://www.aips.nrao.edu}} via the Python wrapper ParselTongue \citep{Parsel}.

\section{Testing the algorithm}
\label{Simuls}

\subsection{Simulating phased-ALMA observations}

We have developed a simulator program to test the performance of the calibration/conversion algorithm implemented in {\em PolConvert}. Our simulation program, {\em PolSimulate}, generates synthetic data under realistic conditions. It takes into account noise from the receivers, signal quantization (amplitude and time), and corruption effects (atmosphere and receivers). The simulator creates a synthetic set of ALMA-only cross-correlations (on a pure linear basis) and the corresponding phased-up data streams, which are cross-correlated with simulated streams from a VLBI station with a circular-feed receiver. We then calibrate the ALMA-only visibilities following the usual ALMA reduction procedures to derive the gain matrices of all the ALMA antennas. Finally, we let {\em PolConvert} compute the Jones matrix for the phased ALMA, apply it to the VLBI fringes, and rewrite them on a pure circular basis. {\em PolSimulate} generates synthetic data following the steps summarized in the following lines: 

\begin{enumerate}

\item  Two random streams are created that simulate the signal from a source on R/L polarization basis. These streams can carry information on all  four of the Stokes parameters, I, Q, U, and V.

\item The R and L streams are phase-rotated according to the parallactic angle of each scan in the simulation. The observations consist of a first scan of an amplitude calibrator (in this case the user can define Stokes I and V) and several scans of a polarization calibrator (the user can define all four Stokes parameters), equally spaced in time between transit and a maximum hour angle. The number of scans and hour-angle coverage are given by the user in a configuration file.

\item Random noise (of amplitude determined by the configuration parameter $T_{ant}/T_{sys}$) is added to the R/L streams. 

\item The noise-free R/L streams are converted to linear basis (alt-az mounts assumed). The signal of each ALMA antenna is then computed as this signal plus uncorrelated noise (of amplitude similar to that in the previous step).

\item The signal from each ALMA antenna is corrupted with gain, bandpass, and leakage. These corrections are different for each antenna and polarization.

\item The (X/Y) signals of the ALMA antennas are cross-correlated, and the result is saved in a measurement set (the data format used by CASA).

\item The phased signal (i.e.,  the sum of the signals from all the ALMA antennas) is cross-correlated with the signal resulting from step 3. The result is saved in a FITS-IDI file.

\end{enumerate}

Quantization to 2-bit and Nyquist time sampling are used when computing all the cross-correlations. The gains of each ALMA antenna as a function of frequency are randomized sinusoidal functions

\begin{equation}
G(\nu) = G_0\,\sin(G_1\,\nu + G_2)
\end{equation}

\noindent with random amplitude ($G_0$), period ($G_1$), and phase ($G_2$). The maximum  values of amplitude and period allowed in these random distributions are provided by the user in the configuration file of {\em PolSimulate}. 

After the simulation, we calibrate the measurement set generated in step 6 using the standard calibration approach. The first scan is used to calibrate the bandpass and the gain (see step 2), whereas the rest of the observations are used to derive the cross-delay, X-Y phase offset, and D-terms.

We have simulated a phased-ALMA mm-VLBI dataset with the following parameters:

\begin{itemize}
\item Observing frequency: $\nu = 100$\,GHz;
\item Bandwidth: $\Delta\nu = 100$\,MHz;
\item Number of ALMA antennas: $N_{ant} = 10$;
\item Receiver noise: $T_{ant}/T_{sys} = 0.1$;
\item Integration time: $t_{int} = 1$\,s;
\item Scan duration: $t_{sc} = 25$\,s;
\item Correlation channels: $N_{chan} = 64$;
\item Number of scans: $N_{sc} = 5$;
\item Declination of sources: $\delta = -60^{\circ}$;
\item Maximum hour angle: $H_{max} = 6$\,h;
\item Maximum leakage (amp \& phase): $D_{max} = 3$\%;
\item Maximum bandpass (amplitude): $B^A_{max} = 10$\%;
\item Maximum bandpass (phase): $B^P_{max} = 20^{\circ}$;
\item Maximum X-Y amplitude ratio: $(G_x/G_y)_{max} = 40$\%;
\item Minimum period $G_1$: 0.25\,cycles/$\Delta\nu$;
\item Maximum period $G_1$: 2.00\,cycles/$\Delta\nu$;
\item Amplitude calib. (Jy): $I=1.0$, $Q=U=V=0.0$;
\item Polarization calib. (Jy): $I=1.0$, $Q=0.1$, $U=V=0.0$.
\end{itemize}

\begin{figure*}[ht!]
\centering
\includegraphics[width=18cm]{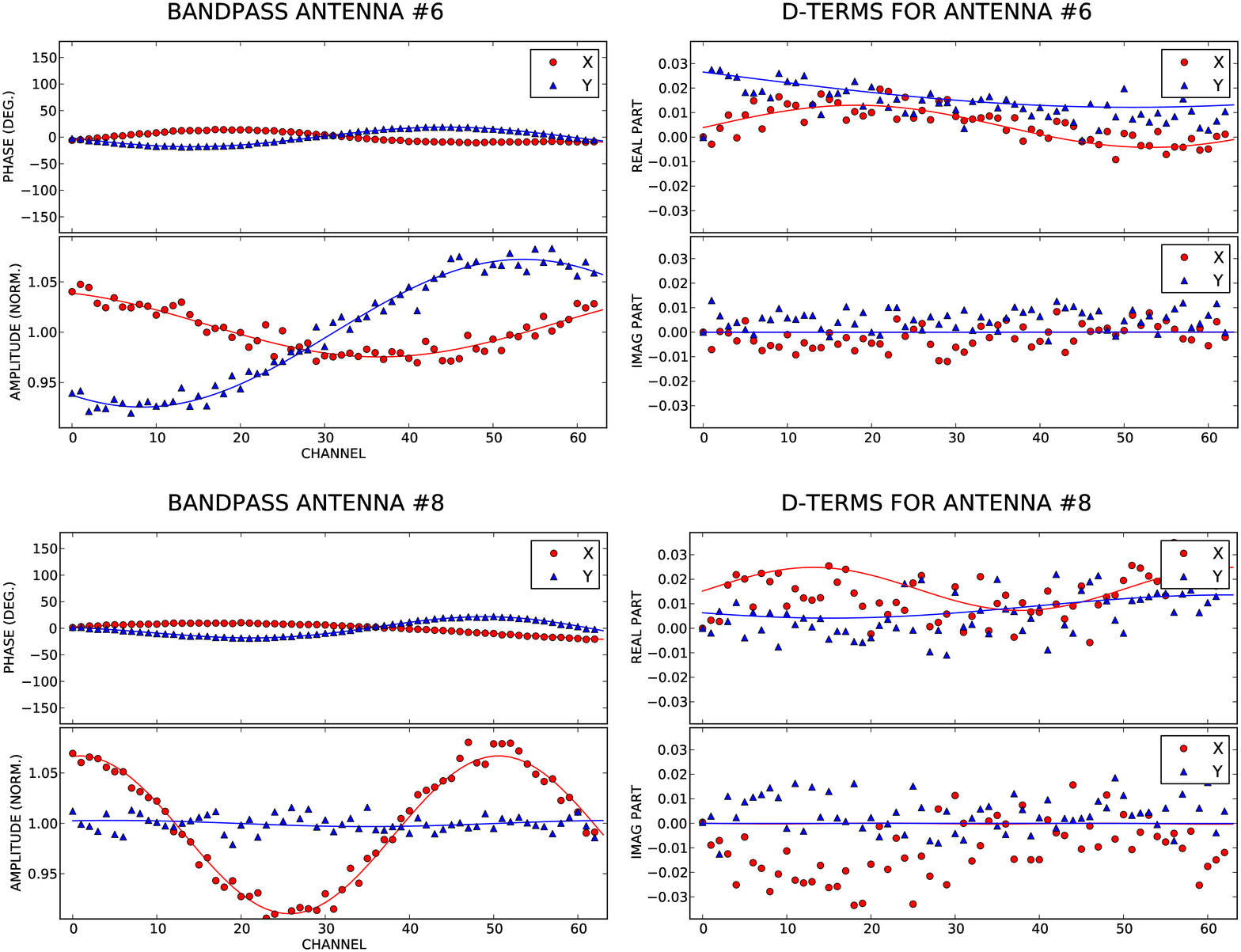}
\caption{Bandpass and leakage gains for two of the simulated antennas. Lines are the gains used in the simulation; points are the gain values derived by CASA.}
\label{GainFig}
\end{figure*}

The magnitudes used for the gains are educated guesses based on our experience in the calibration of real ALMA data. Using these parameters, we obtained a measurement set (with the ALMA-only visibilities) and a FITS-IDI file (with the VLBI visibilities). Some of the gains for the ALMA antennas are shown in Fig. \ref{GainFig}. In all cases, the CASA estimates are very similar to the inputs used in the simulation. This gives us confidence on the reliability of {\em PolSimulate}. We note that the Stokes parameters of the polarization calibrator were left free in the calibration to make the calibration as realistic as possible. The results estimated by CASA when solving for the X-Y phase (i.e., running {\texttt gaincal} in mode ``XYf+QU'') are $Q = 0.0763$ and $U = 0.0089$, not far from the values used in the simulation (i.e., $Q = 0.1$ and $U = 0.0$).

\begin{figure*}[ht!]
\centering
\includegraphics[width=19cm]{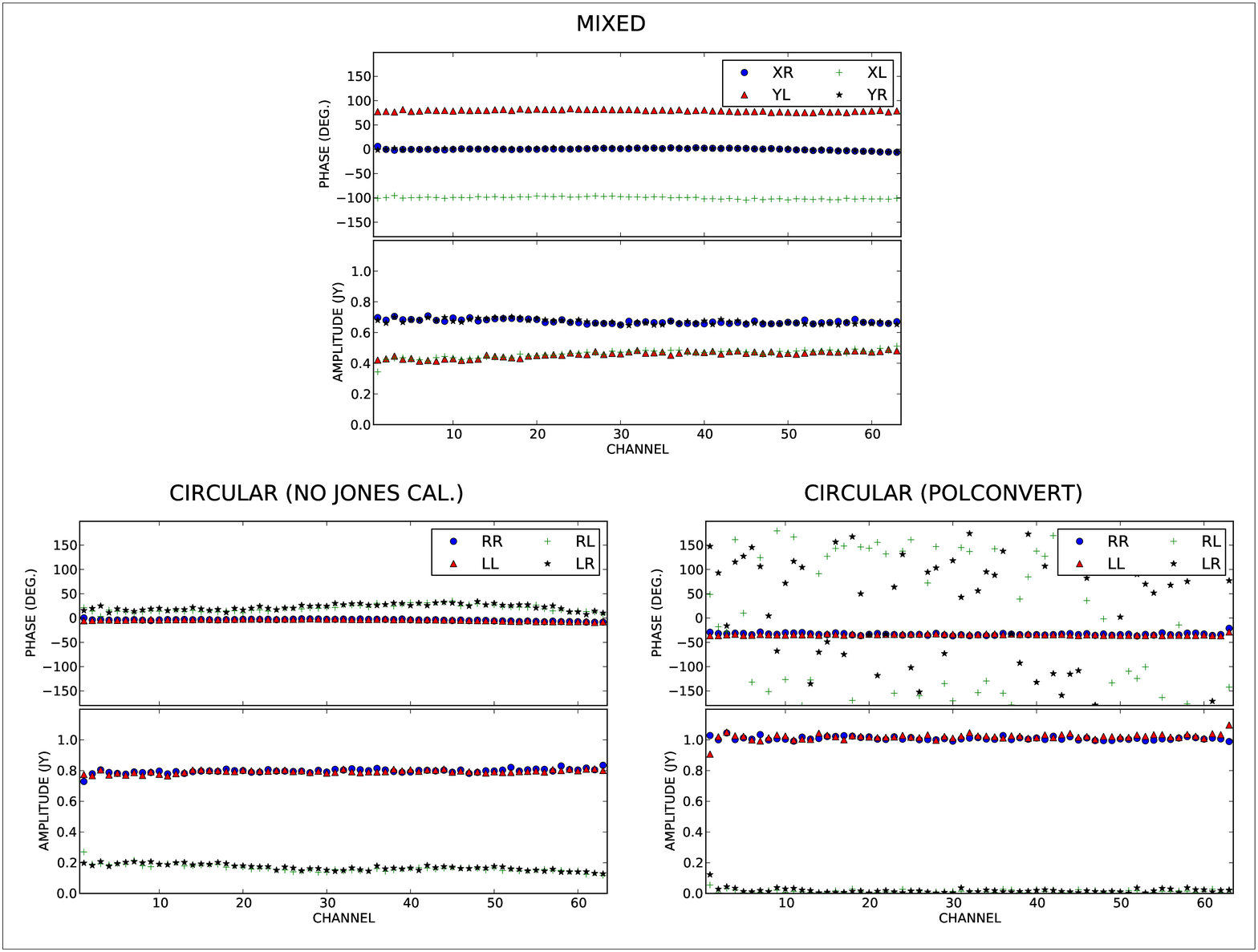}
\caption{VLBI visibility of the unpolarized amplitude calibrator. Top,  mixed-polarization basis. Bottom, after conversion to  pure circular basis. Bottom left, with no corrections based on the gains derived from the cross-correlation of the ALMA antennas. Bottom right, after  full calibration and conversion with {\em PolConvert}.}
\label{VLBIFig}
\end{figure*}

We show the visibilities of the amplitude calibrator in Fig. \ref{VLBIFig}. The visibilities on a mixed basis (i.e., linear to circular) are shown at the top of the figure. Strong bandpass effects and phase offsets can be clearly seen. At bottom left, we show the visibilities converted in a circular basis, but without applying any calibration matrix. This figure illustrates the quality that we would obtain if the polarization conversion was applied either at the recording or at the correlation stage of the VLBI observations (i.e., {\em before} knowing the true gains of each individual ALMA antenna). There is a large polarization leakage left in the cross-hand correlations (about 10\% of the source flux density). Finally, at bottom right we show the visibilities resulting from the full calibration and conversion using {\em PolConvert}. All bandpass and leakage artifacts disappear and the resulting visibility is fully corrected; however, a residual phase in the parallel-hand correlations can  be clearly seen. This phase cannot be calibrated with the Jones matrix derived from the ALMA-only correlations since it is related to the time arrival of the signal to ALMA with respect to the other VLBI station. Obviously, such a time delay is not observable by ALMA alone and is hence transparent to the {\em PolConvert} calibration. However, this phase can be completely corrected with an ordinary fringe-fitting calibration after running {\em PolConvert}.

\begin{figure*}[th!]
\centering
\includegraphics[width=19cm]{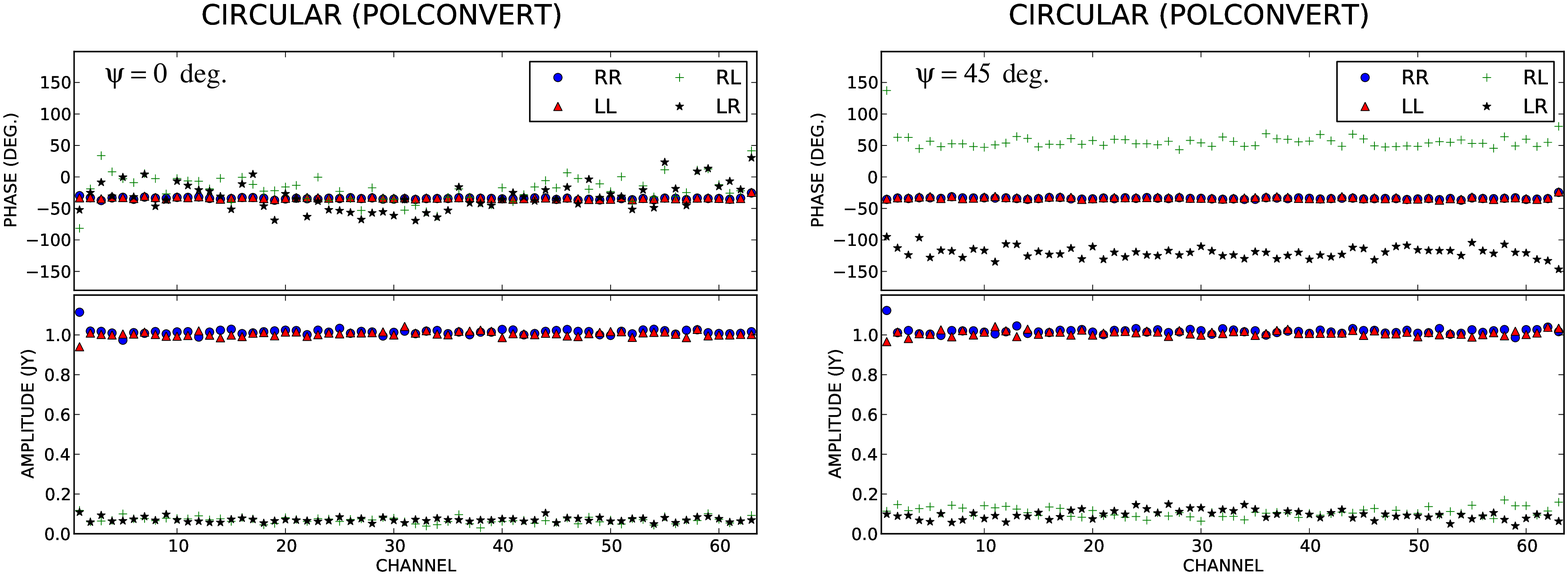}
\caption{VLBI visibility of the polarization calibrator ($Q=0.1; U=V=0.0$) after conversion to a pure circular basis. Left, for a parallactic angle of $0^{\circ}$; right, for a parallactic angle of $45^{\circ}$.}
\label{VLBIFig2}
\end{figure*}

\begin{figure*}[th!]
\centering
\includegraphics[width=15cm]{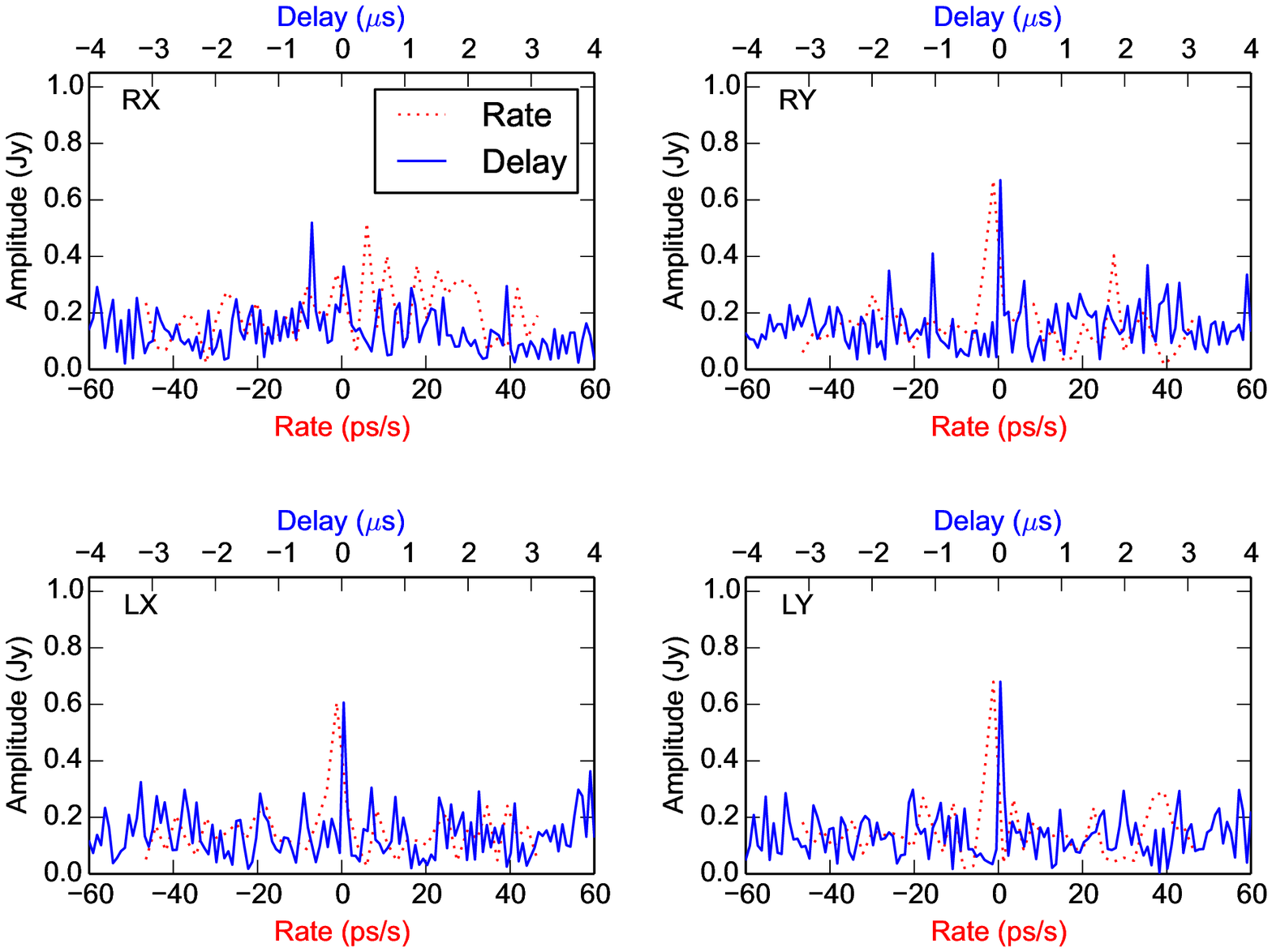}
\caption{VLBI visibility matrix at 86\,GHz between Onsala and Effelsberg  in mixed polarization. Only one sub-band is shown.}

\label{ONEBFig1}
\end{figure*}

\begin{figure*}[th!]
\centering
\includegraphics[width=15cm]{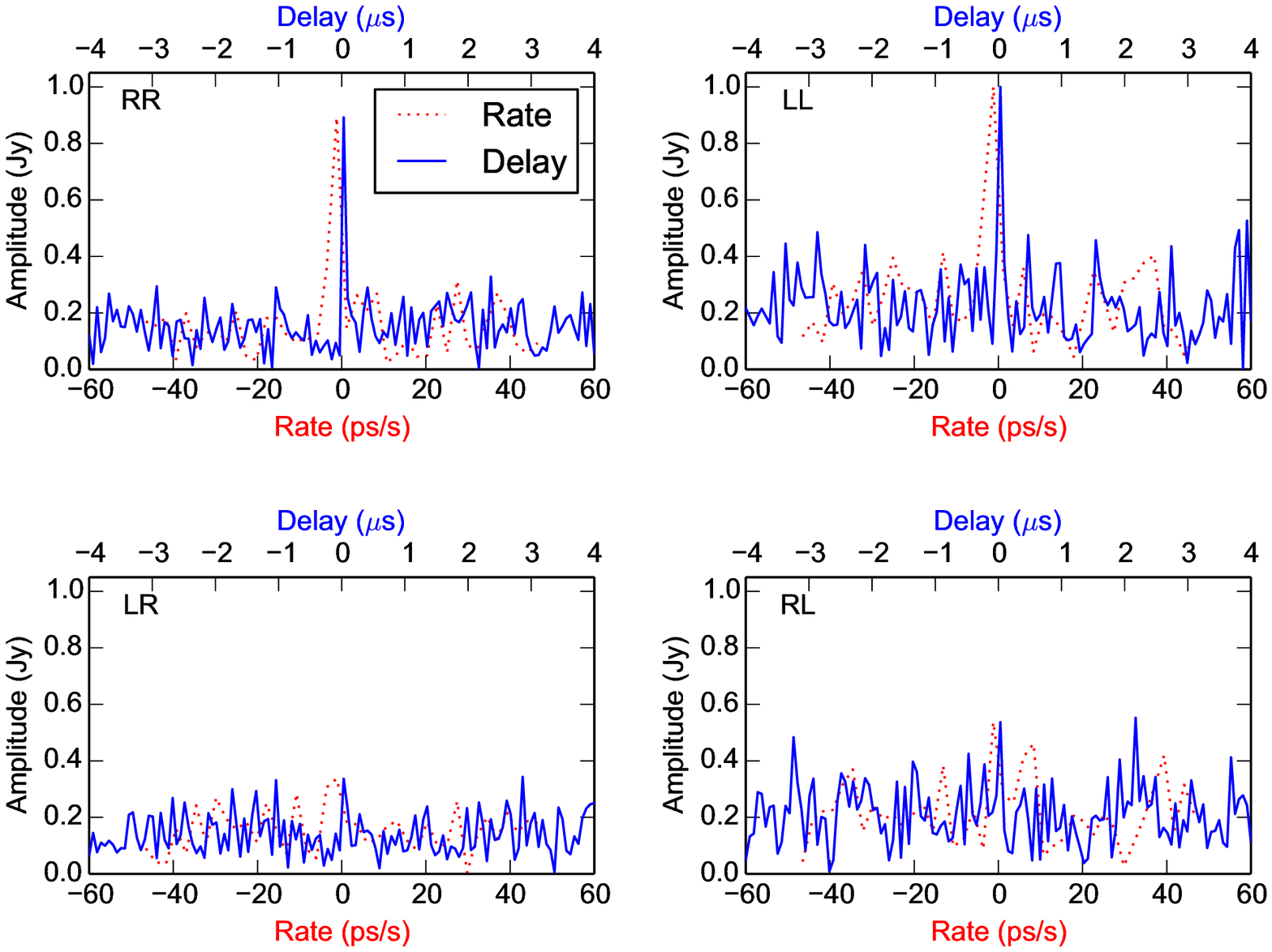}
\caption{VLBI visibility matrix at 86\,GHz between Onsala and Effelsberg in circular polarization. Only one sub-band is shown.} 
\label{ONEBFig2}
\end{figure*}

In Fig. \ref{VLBIFig2}, we show two scans of the polarization calibrator. On the left we show the visibility when the source is in transit (i.e., parallactic angle $\psi = 0$). The expected signal in RL and LR is recovered (i.e., the effect of Q translates into equal and real-valued correlations RL and LR). On the  right, we show the results for the second scan of the polarization calibrator when the parallactic angle is close to $\psi = 45^{\circ}$. Here, the original signal in Q is put into U (in the receiver's frame of the antennas), so now RL and LR become purely imaginary and one is the complex conjugate of the other (RL with positive phase). This is exactly the behaviour that we see in the visibility calibrated and converted with {\em PolConvert}. The phases of RL and LR are equally spaced with respect to those of RR and LL.

\section{Real VLBI observations}
\label{RealData}

On 22 May 2014, we performed a fringe-test VLBI observing session at 86\,GHz using the Onsala 20\,m telescope and the Effelsberg radio telescope. The recording rate was set to 1\,Gbps, synthesizing a bandwidth of 256\,MHz at each polarization channel (8 sub-bands of 32\,MHz; two polarizations).

During part of this session, the staff at Onsala removed the quarter-wave plate at the receiver's front end, which  sent the sky radiation directly into the receiving dipoles on a linear basis. Effelsberg continued data recording with no changes in its receiver, which means that  during that time window Onsala was recording the signal in linear (X/Y) basis, while Effelsberg was recording in circular (R/L) basis. Unfortunately, only 10 seconds of integration time could be used in the correlation owing  to problems in the lock system of the local oscilator at Onsala. During these 10 seconds, we were observing source OJ\,287. a source with a negligible linear polarization. We cross-correlated the data streams from Effelsberg and Onsala with version 2.3 of the program DiFX \citep{DIFX} using a frequency sampling of 128 channels per sub-band and an integration time of 0.25 seconds.

In Fig. \ref{ONEBFig1}, we show the amplitude fringes of an example sub-band of this observation (the second one) in delay-rate space and mixed-polarization basis. Only weak detections are found in the correlation products. In Fig. \ref{ONEBFig2}, we show the fringes after applying the algorithm described in Sect. \ref{SingleStat}. This algorithm was applied to the raw data with no prior global fringe-fitting calibration, because it is not possible to apply it in a standard way to the mixed-polarization visibilities. This is the reason why we could not build the multiband delay to increase the signal-to-noise ratio, S/N, of the detections when we applied the calibration algorithm described in Sect. \ref{SingleStat}. 

Since there is only one baseline in these observations, we had to assume that the amplitude of the cross-polarization gain ratio was unity at Effelsberg. This assumption may have slightly degraded the calibration. If our array had had at least four antennas, we could have derived the true amplitude gain ratios for all the stations without any extra assumption. 

Given the low S/N of our fringes due to the very short integration time, we had to derive the gain ratio for Onsala by averaging over all the frequency channels of the sub-band. Thus, unfortunately, we could not test the performance of the algorithm to calibrate the cross-polarization bandpass effects in the station with linear-feed receivers. In any case, the fringes after the calibration and conversion are very clear in the parallel-hand correlations (RR and LL). The cross-hand correlations (RL and LR) do not show detections, as  should be the case for an unpolarized source.

\section{Conclusions}
\label{Summary}

Interferometric observations using elements with receivers of different polarization configurations may become common in the near future. The case of the phased-ALMA project and its use in VLBI observations is a clear example. ALMA will observe in a linear polarization basis, whereas most of the other VLBI stations will observe in a circular polarization basis. In addition, future VLBI observations at very high data rates (hence wide bandwidths) may imply that it will not be optimum to use quarter-wave plates at some stations, which would degrade the polarimetry quality of their observations. In such cases, these stations will also have to record their signals in linear polarization basis.

We have studied the problem of the calibration and proper polarimetry conversion of interferometric observations performed using mixed polarization  (i.e., some antennas using linear-feed receivers and the rest using circular-feed receivers). We have developed {\em PolConvert}, a software for the calibration/conversion of VLBI mixed-polarization visibilities where the linear-feed station is a phased array. This program is especially designed for its use with ALMA mm-VLBI visibilities. We have successfully tested {\em PolConvert} with realistic synthetic data. We have also developed an algorithm to calibrate and convert visibilities where the linear-feed station is a single dish. This algorithm can also be used to derive the cross-polarization phase offset of the reference antenna in the case of a phased array, if the parallactic angle coverage of the observations is not large enough to ensure a proper full-polarization calibration. We have tested this algorithm with real state-of-the-art VLBI observations at 86\,GHz, performed in mixed polarization, between the Onsala and the Effelsberg radio telescopes. We have shown that it is possible to fully calibrate and convert the mixed-polarization VLBI visibilities, related either to the phased ALMA or to single stations with linear-feed receivers, using our calibration approaches.

\end{document}